\documentclass[a4paper,11pt]{article}
\usepackage{ascmac}
\usepackage{amsthm}
\usepackage[ruled,vlined]{algorithm2e}
\usepackage{amsmath,amssymb}
\usepackage{type1cm}
\usepackage{comment}
\usepackage{here}
\usepackage{float}
\theoremstyle{plain}
\newtheorem{theorem}{Theorem}[section]

\newtheorem{proposition}{Proposition}[section]

\theoremstyle{definition}

\numberwithin{equation}{section}
\newcommand{\msc}[1]{\mbox{{\sc #1}}}
\newcommand{\BQED}{\hfill \hbox{\rule{8pt}{8pt}}}

\newcommand{\malg}{\msc{alg}}
\newcommand{\mperm}{\msc{perm}}

\newcommand{\mopt}{\msc{opt}}
\newcommand{\moff}{\msc{off}}

\newcommand{\ofal}[2]{\mathrm{OFAL}(#1,#2)}
\newcommand{\ofaleq}[2]{\mathrm{OFAL}_{eq}(#1,#2)}
\newcommand{\mc}[1]{\mathcal{#1}}
\title{{\bf {\Large A Lower Bound on the Competitive Ratio of the Permutation Algorithm for Online Facility Assignment 
on a Line}}}

\author{Tsubasa Harada
\thanks{Department of Mathematical and Computing Science,
Tokyo Institute of Technology, 2-12-1 Ookayama, Meguro-ku, 
Tokyo 152-8550, Japan. Email: harada.t.ak@m.titech.ac.jp.}}
\date{}
\begin{document}
\maketitle

\noindent {\sf Abstract:}  
In the online facility assignment on a line (OFAL) with
a set $S$ of $k$ servers and a capacity $c:S\to\mathbb{N}$, 
each server $s\in S$ with a capacity $c(s)$
is placed on a line
and a request arrives on a line one-by-one.
The task of an online algorithm is to 
irrevocably assign a current request to one of the servers with vacancies 
before the next request arrives.
An algorithm can assign up to $c(s)$ requests to each server $s\in S$.

In this paper, we show that the competitive ratio of the permutation algorithm \cite{KalP1993,ARK2020} is at least $k+1$ for OFAL where the servers are evenly placed on a line.
This disproves the result that the permutation algorithm is $k$-competitive
by Ahmed et al. \cite{ARK2020}.
\medskip\\
{\sf Key Words:} Online algorithm, 
Competitive analysis,
Online facility assignment,
Online transportation problem,
Online facility assignment on a line.

\section{Introduction} \label{sec-intro}

\subsection{Background} \label{subsec-background}

The \textit{online facility assignment} (OFA)
or \textit{online transportation} problem
was introduced by
Kalyanasundaram and Pruhs \cite{KalP1995}.
In this problem,
an online algorithm is given a set $S$
of $k$ \textit{servers} and a \textit{capacity} $c:S\to\mathbb{N}$.
Then, the algorithm receives $n$ \textit{requests}
one-by-one in an online fashion.
The task of an online algorithm is to assign each request immediately
to one of the $k$ servers.
Note that the number of requests $n$ is at most
the sum of each server's capacity, i.e.,
$n\leq\sum_{s\in S}c(s)$.
The maximum number of requests that can be
assigned to a server $s\in S$ is $c(s)$, and
the assignment cannot be changed later
once it has been decided.
The cost of assigning a request to a server is
determined by the distance between them.
The goal of the problem is to
minimize the sum of the costs of matching $n$ requests.

A line is considered to be one of the most interesting
metric space for this type of problem for decades
\cite{KalP1998, KN2003, ABNPS2014, AFT2018, ARK2020, IMS2021, HIM2023}.
We refer to a variant of OFA as \textit{OFA on a line} (OFAL)
where all servers and requests are placed on a line. 

Ahmed et al. \cite{ARK2020} dealt with classical competitive analysis for OFAL
under the assumption that the servers are evenly placed
and each server has the same capacity.
We refer to the setting as OFAL$_{eq}$.
Ahmed et al. \cite{ARK2020} showed (with rough proofs) that
the natural greedy algorithm is $4k$-competitive 
and the \textit{permutation} algorithm
\footnote{
Ahmed et al. call this algorithm \textit{Optimal-fill}.
The ``optimal-fill'' algorithm is essentially the same as the permutation algorithm
that has already been proposed by Kalyanasundaram and Pruhs \cite{KalP1993}.
}
is $k$-competitive.
On the other hand, Itoh et al. \cite{IMS2021} analyzed the competitive 
ratio for OFAL$_{eq}$ with small $k\geq 2$.
They showed that (i) for $k=2$, the greedy algorithm is 3-competitive
and best possible, and 
(ii) for $k=3$, $4 $, and $5$, the competitive ratio of any algorithm is at least 
$1+\sqrt{6}>3.449$, $\frac{4+\sqrt{73}}{3}>4.181$, and 
$\frac{13}{3}>4.333$, respectively.

However, when $k=2$, there is a discrepancy between
Ahmed et al.'s result (R1) that the permutation algorithm is $2$-competitive \cite{ARK2020} and Itoh et al.'s result (R2) that the competitive ratio of any algorithm for OFAL$_{eq}$ is at least 3 \cite{IMS2021}.
There has been no research that resolve the contradiction between (R1) and (R2).

\subsection{Our contribution} \label{subsec-contribution}

In this paper, we will show that the competitive ratio of the permutation algorithm
\cite{KalP1993} is in fact at least $k+1$ for OFAL$_{eq}$ (Theorem \ref{thm-k+1}).
This result disproves the claim that the permutation algorithm is $k$-competitive reported by Ahmed et al. \cite{ARK2020} and resolves the contradiction between (R1) and (R2).

\begin{theorem}
\label{thm-k+1}
For $\mathrm{OFAL}_{eq}$,
the competitive ratio of the permutation algorithm is at least $k+1$.
\end{theorem}

\section{Preliminary} \label{sec-preliminary}
%
\subsection{Online Facility Assignment on a Line} \label{subsec-ofal}
%
We define the online facility assignment problem on a line with
$k$ servers and a capacity $c$ which is denoted by $\ofal{k}{c}$.
Let $S=\{s_1, \ldots ,s_k\}\subseteq\mathbb{R}$ be the set of $k$ servers,
$\sigma=r_1\ldots r_n$ ($r_i\in\mathbb{R}$ for $1\leq i\leq n\leq\sum_{s\in S}c(s)$) be a request sequence and $c:S\to\mathbb{N}$ be a capacity.
We can assign up to $c(s)$ requests to each server $s\in S$.

The set $S$ is given to an online algorithm in advance, while requests are given 
one-by-one from $r_{1}$ to $r_{n}$. At any time of the execution of an algorithm, 
a server is called \textit{free}
if the number of requests assigned to it is less 
than its capacity, and \textit{full} otherwise.
When a request $r_{i}$ is revealed,
an online algorithm must assign $r_{i}$ to one of the free servers.
If $r_{i}$ is assigned to $s_{j}$, the pair $(r_{i},s_{j})$
is added to the current matching and the cost 
$|r_{i}-s_{j}|$ is incurred for this pair.
Once the pair $(r_{i},s_{j})$ has been added, an algorithm cannot
remove this pair later.,
The cost of the assignment is the sum of the costs of all the pairs contained in it.
The goal of online algorithms is to minimize the cost of the final matching. 

In this paper, we only deal with the case where all servers are evenly placed
on a line.
Without loss of generality, we assume that
$S=\{s_1,\ldots ,s_k\}=\{0,\ldots ,k-1\}$.
For the case where each server has the same capacity, i.e., $c(s)=\ell$ for each $s\in S$,
we use $\ofaleq{k}{\ell}$ to denote that problem.

\subsection{Notations and Terminologies} 
\label{subsec-notation}
%
For an (online/offline) algorithm {\sc alg} for $\ofaleq{k}{\ell}$ 
and a request sequence 
$\sigma=r_{1}\cdots r_{n}$,  
we use $\msc{alg}(r_{i};\sigma)$ to denote 
the cost of $\malg$ incurred to assign $r_{i}$ when 
{\sc alg} processes $\sigma$.  
For a subsequence $\tau=r_{i_{1}} \cdots r_{i_{m}}$ of 
$\sigma$, we use $\msc{alg}(\tau;\sigma)$ to denote 
the total cost of $\malg$ incurred to assign each $r_{i_{h}}$ 
to the server $s_{\rm alg}(r_{i_{h}};\sigma)$, i.e., 
\[
\msc{alg}(\tau;\sigma)=\sum_{h=1}^{m} \msc{alg}(r_{i_{h}};\sigma).
\]
When $\tau=\sigma$, we simply write $\msc{alg}(\sigma)$ instead of 
$\msc{alg}(\sigma;\sigma)$. 

We use {\sc opt} to denote the optimal {\it offline\/} algorithm, i.e., 
{\sc opt} knows the entire sequence $\sigma=r_{1}\cdots r_{n}$ 
in advance and {\it minimizes\/} the total cost 
to assign all requests from $r_1$ to $r_n$. 

To evaluate the performance of an online algorithm {\sc alg}, 
we use the (strict) competitive ratio. 
We say that {\sc alg} is $\alpha$-competitive for $\ofaleq{k}{\ell}$ if 
$\msc{alg}(\sigma) \leq \alpha \cdot \msc{opt}(\sigma)$ for 
any request sequence $\sigma$ for $\ofaleq{k}{\ell}$. 
For $\ofaleq{k}{\ell}$, the competitive ratio ${\cal R}_{k,\ell}(\msc{alg})$ of {\sc alg}
is defined to be the infimum of $\alpha\geq 1$ such that {\sc alg} is $\alpha$-competitive for $\ofaleq{k}{\ell}$, i.e., 
\[
{\cal R}_{k,\ell}(\msc{alg}) = \inf \{\alpha \geq 1: \mbox{{\sc alg} is $\alpha$-competitive for $\ofaleq{k}{\ell}$}\}. 
\]
If $\malg$ is not $\alpha$-competitive for any $\alpha\geq 1$, then we define ${\cal R}_{k,\ell}(\msc{alg})$ to be $\infty$.

\subsection{Permutation Algorithm} 
\label{subsec-defperm}

In this subsection, we define the permutation algorithm
(denoted by $\mperm$) for $\ofal{k}{c}$ \cite{KalP1993, ARK2020}.
Fix any request sequence $\sigma=r_1\ldots r_n$ for $\ofal{k}{c}$ arbitrarily
and let $M_i=\{(r_j, s^{(j)})\}_{j=1}^i$ be an optimal assignment of $r_1\ldots r_i$.
For $s\in S$, we use $n_{i}(s)$ to denote the number of $j$ such that $s^{(j)}=s$,
i.e., the number of requests assigned to $s$ in $M_i$.
It is known that there is a sequence of optimal assignments $\{M_i\}_{i=1}^n$
that satisfies the following property \cite{KalP1993}: for each $i=1,\ldots ,n$,
\begin{center}
there is a unique $s_{1}^{*}$ such that $n_{i}(s_{i}^{*})=n_{i-1}(s_{i}^{*})+1$ 
and $n_{i}(s) = n_{i-1}(s)$ for any $s \neq s_{i}^{*}$.~~~(2.1)
\end{center}
Note that $n_0(s)=0$ for each $s\in S$.
Since this property holds for optimal assignments,
we can define $\mperm$ as follows:
when an $i$-th request $r_i$ is revealed, $\mperm$ computes
the optimal assignment $M_i$ of $r_1\ldots r_i$ that satisfies (2.1)
and then assigns $r_i$ to $s_i^*$.

\section{The Proof of Theorem \ref{thm-k+1}} 
\label{sec-lbperm}

In this section, we will prove Theorem \ref{thm-k+1}.
To begin with, we show the following proposition.

\begin{proposition}
\label{prop-ell1}
Let $\malg$ be any online algorithm for $\ofaleq{k}{\ell}$
such that $\mc{R}_{k,1}(\malg)<\infty$.
Then, $\mc{R}_{k,\ell}(\malg)\geq\mc{R}_{k,1}(\malg)$
for any $\ell\in\mathbb{N}$.
\end{proposition}

\noindent \textbf{Proof:}
If there exists a request sequence $\sigma$ for $\ofaleq{k}{\ell}$
such that $\malg(\sigma)>0$ and $\mopt(\sigma)=0$, then
$\malg(\sigma)\leq \alpha\cdot\mopt(\sigma)$ does not hold
for any $\alpha\geq1$, i.e., $\mc{R}_{k,\ell}(\malg)=\infty$.
Then, we have $\mc{R}_{k,1}(\malg)<\mc{R}_{k,\ell}(\malg)$.
In the rest of the proof, consider the case where
$\malg(\sigma)=0$ if $\mopt(\sigma)=0$ for any $\sigma$. 

Let $\mc{R}_{k,1}(\malg)$ be $\alpha$.
By the definition of the competitive ratio,
there exists a request sequence $\sigma'=r'_1,\ldots ,r'_k$
such that $\malg(\sigma')\geq \alpha\cdot\mopt(\sigma')$.
Let us define a request sequence $\sigma=r_1,\ldots ,r_{k\ell}$ for $\ofaleq{k}{\ell}$
such that $\malg(\sigma)\geq \alpha\cdot\mopt(\sigma)$ as follows:
for each $i=1,\ldots ,k$, we first give $\ell -1$ requests on $s_i$ and then
give $k$ requests from $r'_1$ to $r'_k$.
Since $\mopt(r_1\ldots r_{k(\ell-1)};\sigma)=0$,
we have $\malg(r_1\ldots r_{k(\ell-1)};\sigma)=0$.
Hence, both $\malg$ and $\mopt$ can assign only one request to each server after processing $r_{k(\ell-1)}$. Thus, we obtain
\begin{align*}
\malg(\sigma)=\malg(\sigma')\geq \alpha\cdot\mopt(\sigma')=\alpha\cdot\mopt(\sigma).
\end{align*}
and this implies that $\mc{R}_{k,\ell}(\malg)\geq \alpha=\mc{R}_{k,1}(\malg)$.
\BQED
\vskip.5\baselineskip

Now we are ready to prove Theorem \ref{thm-k+1}.
We will show that there exists a request sequence $\sigma=r_1\ldots r_k$
for $\ofaleq{k}{1}$ such that
\[
\mperm(\sigma)\geq (k+1-\epsilon)\mopt(\sigma)
\]
for any $\epsilon>0$.
By Proposition \ref{prop-ell1}, this implies $\mc{R}_{k,\ell}(\mperm)\geq k+1$.
We consider the following two cases: (1) $k$ is even and (2) $k$ is odd.

\vskip.5\baselineskip
\noindent \textbf{(1) $k$ is even}
\vskip.5\baselineskip

Define $\sigma=r_1\ldots r_k$ as follows: for $j=1,\ldots ,k/2$, let
\begin{align*}
r_{2j-1} &= s_{k/2 + j} - \frac12 - \epsilon_{2j-1} \text{ and}\\
r_{2j} &= s_{k/2 -j+1} - \frac12  +\epsilon_{2j},
\end{align*}
where
\begin{equation}
\label{eq-def-ep}
\epsilon_i = 
\begin{cases}
\frac{\epsilon}{8}\cdot 2^{i-k} & \text{$i<k$,} \\
\frac12 & \text{$i=k$.}
\end{cases}
\end{equation}
Recall that $s_i=i-1$ for $i=1,\ldots ,k$.
By the definition of $\mperm$, we have that for $j=1,\ldots ,k/2$,
$\mperm$ assigns $r_{2j-1}$ to $s_{k/2-j+1}$ and $r_{2j}$ to $s_{k/2+j}$.
In addition, we consider the offline algorithm $\moff$ that assigns $r_{2j-1}$ to $s_{k/2+j}$
and $r_{2j}$ to $s_{k/2-j+1}$.
Then, we have
\begin{align*}
\mperm(\sigma)&= |r_1-s_{k/2}| + |r_2-s_{k/2+1}| + \cdots + |r_{k-1}-s_1| + |r_k-s_k| \\
&\geq \left( \frac12-\epsilon_1 \right) + \left( \frac32-\epsilon_2 \right)
+ \cdots + \left( \frac{2k-3}{2}-\epsilon_{k-1} \right) + (k-1) \\
&= \frac{(k-1)(k+1)}{2}-(\epsilon_1+\cdots +\epsilon_{k-1}) \\
&= \frac{(k-1)(k+1)}{2}-\frac{\epsilon}{8}\left( 2^{1-k}+\cdots +2^{-1} \right) \\
&\geq \frac{(k-1)(k+1)}{2}-\frac{\epsilon}{8}
\end{align*}
and
\begin{align*}
\mopt(\sigma)&\leq \moff(\sigma)\\ 
&= |r_1-s_{k/2+1}| + |r_2-s_{k/2}| + \cdots + |r_{k-1}-s_{k-1}| + |r_k-s_1| \\
&\leq \left( \frac12+\epsilon_1 \right) + \left( \frac12+\epsilon_2 \right)
+ \cdots + \left( \frac{1}{2}+\epsilon_{k-1} \right) + 0 \\
&\leq \frac{k-1}{2}+\frac{\epsilon}{8}.
\end{align*}
Therefore, we finally get

\begin{align*}
\frac{\mperm(\sigma)}{\mopt(\sigma)}&\geq \frac{\mperm(\sigma)}{\moff(\sigma)}
\geq \frac{\frac{(k-1)(k+1)}{2}-\frac{\epsilon}{8}}{\frac{k-1}{2}+\frac{\epsilon}{8}}
= k+1- \frac{(k+2)\frac{\epsilon}{8}}{\frac{k-1}{2}+\frac{\epsilon}{8}} \\
&\geq k+1- \frac{\epsilon}{4}\cdot \frac{k+2}{k-1} \geq k+1-\epsilon .
\end{align*}

\vskip.5\baselineskip
\noindent \textbf{(2) $k$ is odd}
\vskip.5\baselineskip

Define $\sigma=r_1\ldots r_k$ as follows: for $j=1,\ldots ,(k+1)/2$, let
\begin{align*}
r_{2j-1} &= s_{(k-1)/2 + j} + \frac12 - \epsilon_{2j-1} \text{ and}\\
r_{2j} &= s_{(k-1)/2 -j+1} + \frac12  +\epsilon_{2j},
\end{align*}
where we do not apply
the formula $r_{2j} = s_{(k-1)/2 -j+1} + \frac12  +\epsilon_{2j}$
for $j=(k+1)/2$.
Note that
$\{\epsilon_i\}_{i=1}^k$ is defined by the formula (\ref{eq-def-ep}).
By the definition of $\mperm$, for $j=1,\ldots ,(k+1)/2$,
$\mperm$ assigns $r_{2j-1}$ to $s_{(k-1)/2-j+1}$ and $r_{2j}$ to $s_{(k-1)/2+j}$.
In addition, we consider the offline algorithm $\moff$ that assigns $r_{2j-1}$ to $s_{(k-1)/2+j}$
and $r_{2j}$ to $s_{(k-1)/2-j+1}$.
Thus, similarly to the case (1), we can show that

\begin{align*}
\mperm(\sigma) &\geq \frac{(k-1)(k+1)}{2}-\frac{\epsilon}{8}, \\
\mopt(\sigma) & \leq\moff(\sigma)\leq \frac{k-1}{2}+\frac{\epsilon}{8},
\end{align*}
and then $\mperm(\sigma)/\mopt(\sigma)\geq k+1-\epsilon$.
This completes the proof.

\section{Concluding Remarks and Open Problems} \label{sec-conclusion}

In this paper, we showed that the competitive ratio of $\mperm$ is at least $k+1$ for $\ofaleq{k}{\ell}$.
We conjecture that the competitive ratio of $\mperm$ is exactly $k+1$, but we have not yet been able to prove it.
Furthermore, since we found that $\mperm$ is not $k$-competitive, it is possible that the competitive ratio of $\mperm$ depends on a capacity $\ell$.

\bibliography{optmpfs-bib.bib}

\begin{thebibliography}{1}

\bibitem{ARK2020}
Abu~Reyan Ahmed, Md~Saidur Rahman, and Stephen Kobourov.
\newblock Online facility assignment.
\newblock {\em Theoretical Computer Science}, 806:455--467, 2020.

\bibitem{ABNPS2014}
Antonios Antoniadis, Neal Barcelo, Michael Nugent, Kirk Pruhs, and Michele
  Scquizzato.
\newblock A ${o}(n)$-competitive deterministic algorithm for online matching on
  a line.
\newblock In {\em International Workshop on Approximation and Online
  Algorithms}, pages 11--22. Springer, 2014.

\bibitem{AFT2018}
Antonios Antoniadis, Carsten Fischer, and Andreas T{\"o}nnis.
\newblock A collection of lower bounds for online matching on the line.
\newblock In {\em LATIN 2018: Theoretical Informatics: 13th Latin American
  Symposium, Buenos Aires, Argentina, April 16-19, 2018, Proceedings 13}, pages
  52--65. Springer, 2018.

\bibitem{HIM2023}
Tsubasa Harada, Toshiya Itoh, and Shuichi Miyazaki.
\newblock Capacity-insensitive algorithms for online facility assignment
  problems on a line.
\newblock {\em online ready in Discrete Mathematics, Algorithms and
  Applications}, 2023.

\bibitem{IMS2021}
Toshiya Itoh, Shuichi Miyazaki, and Makoto Satake.
\newblock Competitive analysis for two variants of online metric matching
  problem.
\newblock {\em Discrete Mathematics, Algorithms and Applications},
  13(06):2150156, 2021.

\bibitem{KalP1993}
Bala Kalyanasundaram and Kirk Pruhs.
\newblock Online weighted matching.
\newblock {\em Journal of Algorithms}, 14(3):478--488, 1993.

\bibitem{KalP1998}
Bala Kalyanasundaram and Kirk Pruhs.
\newblock On-line network optimization problems.
\newblock {\em Online algorithms: the state of the art}, pages 268--280, 2005.

\bibitem{KalP1995}
Bala Kalyanasundaram and Kirk~R Pruhs.
\newblock The online transportation problem.
\newblock {\em SIAM Journal on Discrete Mathematics}, 13(3):370--383, 2000.

\bibitem{KN2003}
Elias Koutsoupias and Akash Nanavati.
\newblock The online matching problem on a line.
\newblock In {\em International Workshop on Approximation and Online
  Algorithms}, pages 179--191. Springer, 2003.

\end{thebibliography}
\bibliographystyle{plain}

\end{document}